\documentclass[%
 reprint,
showpacs,
 amsmath,amssymb,
 aps,pra,
]{revtex4-1}

\usepackage{graphicx}
\usepackage{dcolumn}
\usepackage{bbold}
\usepackage{amsmath}
\usepackage{amssymb}
\usepackage[normalem]{ulem}

\newcommand{\ie}{\textit{i.e.}}
\newcommand{\md}{\mathrm{d}}
\newcommand{\etal}{\textit{et al.}}
\newcommand{\re}{\mathrm{Re}}
\newcommand{\im}{\mathrm{Im}}

\newcommand{\tin}{\text{in}}
\newcommand{\tout}{\text{out}}
\newcommand{\xzpf}{x_\text{ZPF}}
\newcommand{\tth}{\text{th}}
\newcommand{\opt}{\text{opt}}

\begin{document}


\title{Generating quadrature squeezed light with dissipative optomechanical coupling}

\author{Kenan Qu and G. S. Agarwal}%
\affiliation{%
 Department of Physics, Oklahoma State University, Stillwater, Oklahoma - 74078, USA
}%


\date{\today}

\begin{abstract}
 The recent demonstration of cooling of a macroscopic silicon nitride membrane based on dissipative coupling makes dissipatively coupled optomechanical systems promising candidates for squeezing. We theoretically show that such a system in a cavity on resonance can yield good squeezing which is comparable to that produced by dispersive coupling. We also report the squeezing resulting from the combined effects of dispersive and dissipative couplings; thus the device can be operated in one regime or the other. We derive the maximal frequency and quadrature angles needed to observe squeezing for given optomechanical coupling strengths. We also discuss the effects of temperature on squeezing.
\end{abstract}

\maketitle


\section{Introduction}
The field of cavity optomechanics continues to register significant progress and a comprehensive review has recently appeared~\cite{RMP}. The important developments include cooling of the mirror to its ground state~\cite{cooling1,cooling2,cooling3}, mode splitting~\cite{NMS1,NMS2}, quantum state engineering of the mechanical motion~\cite{loop3,loop4,OMS2006}, and electromagnetically induced transparency (EIT) and its various applications~\cite{EIT1,EIT2,EIT3,EIT4,EIT5,EIA,slowlight1,slowlight2}. More recently, strictly quantum effects like squeezing of the mirror~\cite{APB,sqZoller,Nori,other1,other2,other3,other4,other5} and the cavity field~\cite{sq0,sq1,sq2,sq3,PRX,kenan-NJP,Woolley,Kronwald,squeezing,Vitali,Asjad} as well as generation of entangled photon pairs~\cite{OMSEnt,Lehnert1,Lehnert2} are receiving considerable attention. Ponderomotive squeezing of light~\cite{sq0,sq1,sq2,sq3,PRX,Vitali} using an on-resonance driving laser is one of the most promising ways to generate squeezed light in cavity optomechanics. Safavi-Naeini \etal~\cite{sq3} fabricated a micromechanical cavity resonator from a silicon microchip and they observed the fluctuation spectrum at a level $(4.5\pm0.2)\%$  below the shot-noise limit  despite highly excited thermal state of the mechanical resonator ($10^4$ phonons). Purdy \etal~\cite{PRX} placed a low-mass partially reflective membrane made of silicon nitride in the middle of an optical cavity and pushed the squeezing limit to $32\%$ ($1.7$dB) by cooling the membrane to about $1$mK. Additional ways of producing optical squeezing in optomechanical systems have also been proposed. One example is use of a double-cavity optomechanical system to generate  two-mode squeezed light~\cite{kenan-NJP,Woolley}. Another example~\cite{Kronwald} is generation of quadrature squeezed light using the dissipative nature of the mechanical resonator in a single cavity driven by two differently detuned lasers. In a closely related subject, Lehnert and co-workers reported the experimental realization of entanglement between cavity output photon-photon pairs~\cite{Lehnert1} and entanglement between mechanical motion and microwave fields~\cite{Lehnert2}.

It should be noted that much of the work on cavity optomechanics uses dispersive coupling. However, there are a few studies for dissipative coupling~\cite{diss,diss1,diss2,diss3, diss4, diss5,diss6,dissSumei}---the intrinsic cavity lifetime depends on the mechanical motion. A theoretical analysis of dissipative coupling in cavity optomechanics was reported by Elste \etal~\cite{diss}. They pointed out that the system gives rise to a remarkable quantum noise interference effect which leads to the Fano line shape in the backaction force noise spectra. Experimentally, Li \etal\ ~\cite{diss4} for the first time reported dissipative coupling in a cavity optomechanics system that comprises a microdisk and a vibrating nanomechanical beam waveguide. Based on such a setup, Huang and Agarwal~\cite{dissSumei} proposed a scheme to beat the standard quantum limit (SQL) by irradiation of squeezed light into the cavity. Hammerer and co-workers~\cite{diss1,diss2} concentrated on dissipative coupling by placing an optomechanical membrane inside a Michelson-Sagnac interferometer. This scheme is advantageous in the sense that the dissipative coupling is not due to internal dissipation, but the output photons are detectable. Weiss \etal~\cite{diss3} presented a comprehensive study of dissipative coupling in both the weak- and strong-coupling limits, and they found the parameter regions for amplification of cooling as well as EIT and normal-mode splitting.  Wu \etal~\cite{diss5} experimentally reported the application of torque sensing by using dissipative optomechanical coupling in a photonic crystal split-beam nanocavity.
Very recently, Sawadsky \etal~\cite{combined} demonstrated cooling starting from room temperature to $126$mK based on the combined effects of dissipative and dispersive coupling. This is quite a remarkable development where the couplings can be changed adding flexibility to the operation. Encouraged by the significant cooling in this experiment, we examine the optical squeezing that can be produced in a dissipative optomechanical interaction.

In this paper, we develop analytically the theory of ponderomotive squeezing in cavity optomechanics with dissipative coupling. We show that the squeezing magnitudes with dissipative coupling are comparable to those achieved using dispersive coupling. This squeezing scheme broadens the scope of the quantum study of nonlinear interaction in optomechanics. Our proposal is based on the parameters reported in~\cite{combined}; however, it is not limited to this system and is applicable to any optomechanical systems that can provide combined interactions. This squeezing scheme works in the unresolved-sideband regime, which has advantages in its easier system fabrication requirements. Moreover, this particular parameter regime makes it feasible for obtaining squeezed light with low frequency mechanical oscillators, although thermal phonons are still an issue. We show that the system can generate a $3$dB squeezed field by use of reasonable driving laser powers when the thermal phonon occupancy is as large as $1.5\times10^5$ (the corresponding bath temperature $T=1$K). The effect of a higher bath temperature can be offset by increasing the driving laser power. As a by-product, our theory explains the new instability region for small pump laser red-detunings which was discovered in the experiment~\cite{combined}.

The structure of this paper is organized as follows: In Sec.~\ref{model}, we introduce the Hamiltonian of the optomechanical system with both dispersive and dissipative couplings, and find the input-output relation for the cavity field. In Sec.~\ref{limit}, we provide the analysis of the squeezing effects under purely dissipative coupling. We compare it with the conventional dispersive squeezing and show that they both generate squeezed output with similar magnitudes but in different quadratures. In Sec.~\ref{combined}, we study the effects of the combined coupling on the squeezing and find the optimal quadrature angle for squeezing. We also study the effects of the mechanical mode at finite temperature. In Sec.~\ref{fixed}, we analyze the effective detuning of the driving laser due to the change of cavity resonance frequency, and then show its effect on the squeezing spectra. We present our conclusion in Sec.~\ref{conclusion}.

\section{Model}\label{model}
We consider an optomechanical system in which a mechanical oscillator (frequency $\omega_m$) is coupled to an electromagnetic cavity. We model the cavity mode with the annihilation operator $a$ and the mechanical oscillator with the displacement $x$ and momentum $p$, or with dimensionless operators $Q=x/\xzpf$ and $P=(\hbar\xzpf)p$ where $\xzpf=\sqrt{m\omega_m/\hbar}$ is the mechanical zero-point fluctuation. The mechanical displacements weakly modulate the cavity resonance frequency $\omega_c(Q)$ and damping rate $\kappa(Q)$. We expand them to linear order to get $\omega_c(Q)\cong \omega_c-g_\omega Q$ and $\kappa(Q)\cong \kappa-g_\kappa Q$, where the dispersive coupling constant $g_\omega= (\partial\omega_c/\partial Q)$ and the dissipative coupling constant $g_\kappa= (\partial\kappa/\partial Q)$. In the general cases, the dispersive coupling is larger than the dissipative coupling by a factor of $g_\omega/g_\kappa=\omega_c/\kappa\gg 1$. However, by placing a micro-membrane inside a Michelson-Sagnac interferometer, it has been shown that $g_\kappa$ and $g_\omega$ can be made of the same order.

When the optomechanical system is driven by a strong laser with frequency $\omega_l$ and power $\mathcal{P}$, the Hamiltonian can be written, in the rotating frame, as
\begin{align}\label{01}
  H &= \hbar(\omega_c-\omega_l)a^\dag a + \frac12 \hbar\omega_m(Q^2+P^2) - \hbar g_\omega a^\dag a Q \nonumber \\
  & + i\hbar\sqrt{2\kappa(Q)}[a^\dag(\mathcal{E}_l + a_\tin) - H.c.],
\end{align}
where $\mathcal{E}_l=\sqrt{\frac{\mathcal{P}}{\hbar\omega_l}}$ and $a_\tin$ represents the input vacuum noise. To proceed, we linearize the Hamiltonian following the standard procedure by writing $a=a_s+a_1$,  $P=P_s+P_1$ and $Q=Q_s+Q_1$. The mean values of the steady state can be calculated as
\begin{equation}\label{steady}
  a_s = \frac{\sqrt{2\kappa_s}\mathcal{E}_l}{\kappa_s + i\Delta_s}, \qquad Q_s = (\frac{g_\omega}{\omega_m} - \frac{\Delta_s g_\kappa}{\kappa_s \omega_m}) |a_s|^2,
\end{equation}
and $P_s=0$. Under the effect of driving laser, the mechanical oscillator displacement $Q_s$ modulates both the cavity resonance frequency and the decay rate. Hence we define $\Delta_s=(\omega_c-g_\omega Q_s) -\omega_l$ as the driving laser detuning from the effective cavity resonance frequency; and we define $\kappa_s = \kappa-g_\kappa Q_s$ as the effective cavity decay rate. Both $\Delta_s$ and $\kappa_s$ depend on the power of the driving laser. However, by tuning the driving laser frequency $\omega_l$, one can always cause it to be on resonance with the effective cavity frequency, \ie~$\Delta_s=0$. Under this condition, the effective cavity decay rate is determined by the quadratic equation $\kappa_s^2 - \kappa \kappa_s + 2\mathcal{E}_l^2 g_\omega g_\kappa/\omega_m=0$. In the typical optomechanical systems, the term  $2\mathcal{E}_l^2 g_\omega g_\kappa/\omega_m$ is negligible compared to $\kappa$ and hence $\kappa_s\cong \kappa$. For example with the parameters reported in~\cite{combined}, $2\mathcal{E}_l^2 g_\omega g_\kappa/\omega_m < \kappa /10^3$ when the driving power is below $10$mW.

Then the linearized Hamiltonian takes the form $H = H_0 + H_\text{int} + H_\text{damp}$ and
\begin{align}\label{02}
   H_0 &= \hbar\Delta_s a_1^\dag a_1 + \frac12 \hbar\omega_m(Q_1^2+P_1^2), \nonumber \\
   H_\text{int} &= -\hbar \frac{G_\omega^* a_1 + G_\omega a_1^\dag}{\sqrt{2}} Q_1 - \hbar G_\kappa \frac{a_1 - a_1^\dag}{2\sqrt{2}i} Q_1, \\
   H_\text{damp} &= i\hbar\sqrt{2\kappa_s}(a_1^\dag a_\tin- a_\tin^\dag a_1) - i\hbar \frac{G_\kappa a_\tin^\dag  - G_\kappa^* a_\tin}{2\sqrt{\kappa_s}}Q_1, \nonumber
\end{align}
where $G_{\omega,\kappa}=\sqrt2 a_s g_{\omega,\kappa}$ is the driving field enhanced dispersive (dissipative) coupling constant.  The form of the Hamiltonian (\ref{02}) suggests that it is more intuitive to write the cavity field in terms of its quadratures: $X=(a_1+a_1^\dag)/\sqrt2$, $Y=(a_1-a_1^\dag)/(\sqrt2i)$, and $[X,Y]=i$. In this paper, we are interested in generating squeezing light in the output field. Under the effect of dissipative coupling, the standard input-output relation
\begin{align}\label{03}
  a_\tin+a_\tout &\approx \sqrt{2(\kappa - g_\kappa Q_s)} a_1 - \frac{a_s g_\kappa}{\sqrt{2(\kappa - g_\kappa Q_s)}} Q_1 \nonumber \\
  &= \sqrt{2\kappa_s} a_1 - \frac{g_\kappa a_s}{\sqrt{2\kappa_s}}Q_1,
\end{align}
since $g_\kappa \ll \kappa_s$. In terms of quadratures, the input-output relations are $X_\tin+X_\tout \approx \sqrt{2\kappa_s}X - \frac{G_\kappa}{\sqrt{2\kappa_s}}Q_1$ and $Y_\tin+Y_\tout \approx \sqrt{2\kappa_s}Y$. Hereafter, we first focus on the on resonance driving scenario ($\Delta_s=0$) and then discuss the squeezing effect with detuned driving by relaxing this condition. When $\Delta_s=0$, the coupling strength $G_{\omega,\kappa}$ is real. The dynamics of the system can be described  using the quantum Langevin equations
\begin{gather}
  \frac{1}{\omega_m} \ddot{Q}_1 + \frac{\gamma_m}{\omega_m} \dot{Q}_1 + \omega_m Q_1 = G_\omega X + \frac{G_\kappa}{2} Y - \frac{G_\kappa}{\sqrt{2\kappa_s}} Y_\tin + \xi,  \label{04} \\
  \dot{X} = -\kappa_s X +\frac{G_\kappa}{2} Q_1 + \sqrt{2\kappa_s}X_\tin, \label{05} \\
  \dot{Y} = -\kappa_s Y + G_\omega Q_1 + \sqrt{2\kappa_s}Y_\tin. \label{06}
\end{gather}
Here, $\xi$ models the Brownian noise acting on the mechanical oscillator, and it obeys $\langle \xi(t)\xi(t') \rangle = \gamma_m(2\bar{n}_\tth+1) \delta(t-t')$, where $\bar{n}_\tth$ is the mean phonon occupation number. The correlations for the vacuum field are $2\kappa_s \langle X_\tin(t) X_\tin(t') \rangle = 2\kappa_s \langle Y_\tin(t) Y_\tin(t') \rangle = \kappa_s \delta(t-t')$. In the unresolved-sideband limit $\kappa_s\gg \omega_m\gg \gamma_m$, hence the vacuum noise dominates over the Brownian mechanical noise at low $\bar{n}_\tth$.

\begin{figure}[tb]
 \includegraphics[width=0.35\textwidth]{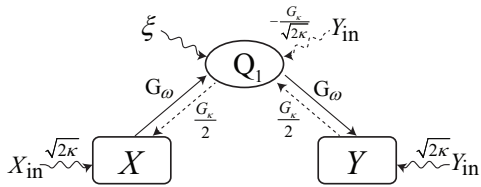}
 \caption{\label{fig2}{The input quantum noises and their coupling relations among different quadratures ($X$, $Y$) of the cavity field and mechanical mode ($Q_1$). The dashed arrows show the noise input and coupling due to dissipative coupling $G_\kappa$. }}
\end{figure}

We illustrate the coupling relations of the quantum noises in the optomechanical system, in Fig.~\ref{fig2}.
The field quadratures are subjected to the vacuum input noise $X_\tin$ and $Y_\tin$. More importantly, we notice that, due to the dissipative coupling $G_\kappa$, the input vacuum noise $Y_\tin$ is also coupled directly to the mechanical motion $Q_1$. At the same time, the form of the interaction Hamiltonian shows that $Q_1$ interacts with the different cavity quadratures at the rates $G_\omega$ and $G_\kappa$. Therefore, $Y_\tin$ is fed into the system through two paths: (i) it directly couples to the cavity field; and (ii) it couples to the mechanical motion $Q_1$ dissipatively and then the optomechanical interaction transfers the noise to the cavity field. These two paths interfere in a coherent manner and lead to the Fano resonance in the cavity field spectrum.

We calculate the output field by combining Eqs.(\ref{03})-(\ref{06}) after taking the Fourier transform, and find
\begin{widetext}
\begin{align}
	(\frac{1}{\chi}-\frac{G_\omega G_\kappa}{\kappa_s-i\omega})(X_\tout - \frac{\kappa_s+ i\omega}{\kappa_s-i\omega} X_\tin) &= \frac{i\omega G_\kappa G_\omega}{(\kappa_s-i\omega)^2} X_\tin -\frac{\omega^2 G_\kappa^2}{2\kappa_s (\kappa_s-i\omega)^2} Y_\tin + \frac{i\omega G_\kappa}{\sqrt{2\kappa_s}(\kappa_s-i\omega)}  \xi,  \label{07} \\
	(\frac{1}{\chi}-\frac{G_\omega G_\kappa}{\kappa_s-i\omega})(Y_\tout - \frac{\kappa_s+ i\omega}{\kappa_s-i\omega} Y_\tin) &= \frac{2\kappa_s G_\omega^2}{(\kappa_s-i\omega)^2} X_\tin + \frac{i\omega G_\kappa G_\omega}{(\kappa_s-i\omega)^2} Y_\tin + \frac{G_\omega \sqrt{2\kappa_s}}{\kappa_s-i\omega} \xi,  \label{08} 
\end{align}
\end{widetext}
where $\chi=\omega_m/(\omega_m^2-\omega^2-i\omega\gamma_m)$ is the mechanical susceptibility. Equations(\ref{07}) and (\ref{08}) describe how the input quantum noises add to the quantum fluctuation of the output fields. Without optomechanical interactions, the output field preserves the input field fluctuations, \ie, $\langle X_\tout^2\rangle = \langle Y_\tout^2\rangle$. As one increases the optomechanical interaction strengths $G_\omega$ and $G_\kappa$, the noises are distributed in a nonlinear manner. The quantum squeezed states are generated when the variance is lower than the that of the coherent state, \ie, $S_\theta=\langle Z_\theta^2 \rangle<1/2$ for a specific quadrature $Z_{\theta\tout} = X_\tout\cos\theta + Y_\tout\sin\theta$.

\section{Squeezing with purely dissipative coupling}\label{limit}
\begin{figure}[htbp]
 \includegraphics[width=0.5\textwidth]{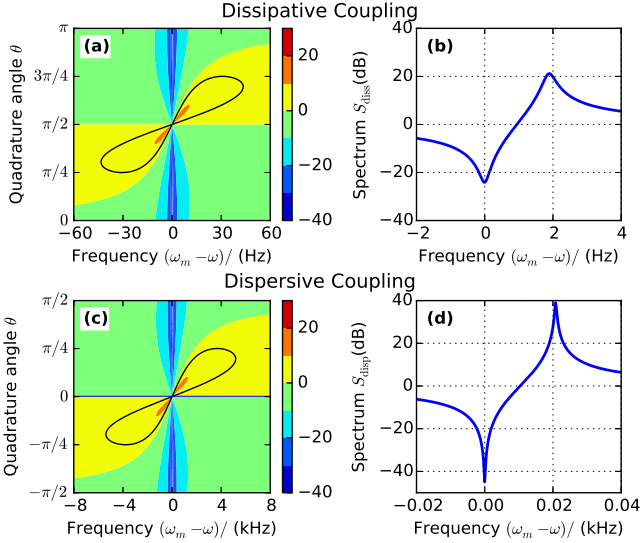}
 \caption{\label{fig3}{(Color online) Comparison of squeezing spectra with purely dissipative coupling (a) and (b) and purely dispersive coupling (c) and (d): The regions inside the black contours in the density plots (a) and (c) show a $3$dB squeezing region and the blue horizontal lines show the optimal quadrature which are plotted in (b) and (d), respectively.  The dissipative coupling strength is $G_\kappa=2\pi\times150$kHz with driving laser power $\mathcal{P}\sim 3.5$W; the dispersive coupling strength is $G_\omega=2\pi\times75$kHz with driving laser power $\mathcal{P}\sim 40$mW. Other parameters are $\kappa_s=2\pi\times1.5$MHz, $\omega_m=2\pi\times136$kHz, $\gamma_m=2\pi\times0.23$Hz, $\Delta_s=0$, and $\bar{n}_\tth=0$.  }}
\end{figure}

The phenomenon of ponderomotive squeezing with purely dissipative coupling can be obtained by setting the dispersive coupling strength $G_\omega=0$ and $\Delta_s=0$, so that $Y_\tout \cong Y_\tin$ and $X_\tout \cong X_\tin - \frac{\chi \omega^2 G_\kappa^2}{2\kappa^3} Y_\tin  +$ mechanical noise. The vacuum input $Y_\tin$ is coupled, not only to $Y_\tout$, but also to $X_\tout$ via the mediated mechanical mode $Q_1$ scaled by the mechanical suspectibility $\chi$ and dissipative coupling strength $G_\kappa$. When one measures the field $Z_{\theta\tout} = X_\tout\cos\theta + Y_\tout\sin\theta$ at $\theta\neq 0\,^\circ$ or $90\,^\circ$, $Y_\tout$ interferes partially with $X_\tout$ since $\chi(\omega)$ is generally complex. The interference leads to squeezed quantum noise. The output squeezing spectrum is
\begin{equation}\label{diss1}
  S_\text{diss} \cong \frac12 + \frac{G_\kappa^2\omega_m^2}{4\kappa_s^3} \big( 2|\chi|^2 \Gamma_\mathrm{diss} \cos^2\theta - \mathrm{Re}\chi \sin2\theta \big),
\end{equation}
where $\Gamma_\mathrm{diss} = G_\kappa^2\omega_m^2/(4\kappa_s^3) + \gamma_m(2\bar{n}_\tth+1)$ is the effective mechanical damping rate. By optimizing $\theta$ and $\chi(\omega)$ we obtain the optimal squeezing magnitude
\begin{equation}\label{diss2}
  S^\opt_\text{diss} = \frac{\gamma_m(\bar{n}_\tth+1)}{G_\kappa^2\omega_m^2/(4\kappa_s^3) + 2\gamma_m(\bar{n}_\tth+1)}.
\end{equation}
The squeezing magnitude can be enhanced by a large effective dissipative optomechanical coupling strength $G_\kappa^2\omega_m^2/(4\kappa_s^3\gamma_m)$ and a low mean phonon occupancy number $\bar{n}_\tth$. The optimal squeezed quadrature angle lies at $\tan\theta^\opt_\text{diss} \cong \sqrt{G_\kappa^2\omega_m^2/(2\kappa_s^3\gamma_m)}$, and $\theta^\opt_\text{diss}$ approaches to $90\,^\circ$ with a large dissipative coupling strength $G_\kappa$.
From the above analysis, we can see that the ponderomotive squeezing relies solely on the interference of two paths of $X_\tin$. One needs to suppress the input noises $Y_\tin$ and $\xi$ by choosing a quadrature angle $\theta^\opt_\text{disp}$ close to $90\,^\circ$. The output field shows anti-squeezing at $\omega=\omega_m$ when $\theta\neq0$. To illustrate the squeezing effect, we plot the output field spectra at different quadratures in Figs.~\ref{fig3}(a) and \ref{fig3}(b)  by numerically solving the quantum Langevin equations (\ref{04})-(\ref{06}). We use the parameters provided by the experiment reported in~\cite{combined}, and the specific values are given in the caption of Fig.~\ref{fig3}. At the angle $\theta^\opt_\text{diss}$, the output spectrum [as shown in (b)] is characterized by a large squeezing of $\sim 20$dB at frequency $\omega\sim\omega_m-2\pi\times2$Hz and anti-squeezing at $\omega=\omega_m$.

In the other limit when dispersive coupling solely governs the optomechanical interaction, \ie, $G_\kappa=0$, Eqs. (\ref{07}) and (\ref{08}) reduce to $X_\tout \cong X_\tin$ and $Y_\tout \cong Y_\tin + (\chi G_\omega^2/\kappa_s) X_\tin + $ mechanical noise. This is the conventional ponderomotive squeezing scheme. It shares a similar noise transformation with that we discussed above. Hence we are able to observe a similar squeezing phenomenon, but the optimal squeezed quadrature is around $\tan\theta^\opt_\text{disp} \cong \sqrt{\kappa_s\gamma_m/(2G_\omega^2)}$, and $\theta^\opt_\text{disp}$ approaches $0$ with a large dispersive coupling strength $G_\omega$.  The output squeezing spectrum is
\begin{equation}\label{disp1}
  S_\text{disp} \cong \frac12 + \frac{G_\omega^2}{\kappa_s} \big( 2|\chi|^2 \Gamma_\mathrm{disp} \sin^2\theta + \mathrm{Re}\chi \sin 2\theta \big),
\end{equation}
where $\Gamma_\mathrm{disp} = G_\omega^2/\kappa_s + \gamma_m(2\bar{n}_\tth+1)$. By optimizing $\theta$ and $\chi(\omega)$ we obtain the optimal squeezing magnitude
\begin{equation}\label{disp2}
  S^\opt_\text{disp} = \frac{\gamma_m(\bar{n}_\tth+1)}{G_\omega^2/\kappa_s + 2\gamma_m(\bar{n}_\tth+1)}.
\end{equation}
This result is identical to the one derived in~\cite{sq0} and has been experimentally demonstrated in \cite{sq3,PRX}.
The optimal output frequency is $(\omega-\omega_m)^2=\Gamma_\mathrm{disp} \gamma_m/2 + \gamma_m^2/4$, which increases with coupling strength $G_\kappa^2$.
We plot the output spectra of dispersive squeezing in Figs.~\ref{fig3}(c) and \ref{fig3}(d), as a comparison with the dissipative squeezing in Figs.~\ref{fig3}(a) and \ref{fig3}(b). The optimal squeezing spectrum has a quadrature angle close to $0$. The optimal squeezing magnitude is shown as $\sim 40$dB, which agrees with Eq.~(\ref{disp2}).  We observe similar output squeezed spectra, but the optimal squeezing magnitude is larger than in Figs.~\ref{fig3}(a) and \ref{fig3}(b).

Physically both the dispersive coupling and the dissipative coupling generate optical squeezing in a similar manner, in the sense that they couple the input noise from one quadrature coherently to the other quadrature. Thus the input vacuum noise couples to the optomechanical system via two paths, as shown in Fig.~\ref{fig2}. These two paths interfere and lead to squeezing. The optimal squeezing exists at different quadrature angles due to the fact that $G_\omega$ couples noise from $X$ to $Y$ and $G_\kappa$ couples noise from $Y$ to $X$ via the mechanical mode.

\section{Squeezing with combined effects of dissipative and dispersive coupling}\label{combined}
In the previous section, we studied squeezing phenomena with purely dispersive coupling or dissipative coupling. One natural question is whether the combined effect of these two coupling regimes could enhance the squeezing. We next study the generation of squeezed state in the presence of both coupling regimes $G_\omega$ and $G_\kappa$.  The output squeezing spectrum is
\begin{align}\label{hybr1}
  S_\text{comb} &\cong \frac{1}{2|D|^2} \left| \frac{1}{\chi}- \frac{G_\kappa G_\omega}{\kappa} \right|^2 
  + \frac{\Gamma_\omega \sin^2\theta + \Gamma_\kappa \cos^2\theta}{2|D|^2} \nonumber \\
  &\qquad + \frac{\Gamma' }{2|D|^2} \mathrm{Re} \bigg[\frac{1}{\chi}- \frac{G_\kappa G_\omega}{\kappa} \bigg] \sin2\theta , \\
  \Gamma_\omega &=  \frac{4G_\omega^4}{\kappa_s^2} + \frac{4G_\omega^2}{\kappa_s}\gamma_m(2\bar{n}_\tth+1) , \nonumber \\
  \Gamma_\kappa &=  \frac{G_\kappa^4\omega^4}{4\kappa_s^6} + \frac{G_\kappa^2\omega^2}{\kappa_s^3}\gamma_m(2\bar{n}_\tth+1) , \nonumber \\
  \Gamma' &=  \frac{2G_\omega^2}{\kappa_s} - \frac{G_\kappa^2\omega^2}{2\kappa_s^3} , \nonumber 
\end{align}
where $D = 1/\chi - G_\omega G_\kappa/(\kappa_s - i\omega)$. The optimal squeezing quadrature angle  $\tan2\theta^\opt_\text{hybr} = (\Gamma_\omega-\Gamma_\kappa)/(2\Gamma')$ and the squeezing magnitude can be found to a rough approximation
\begin{equation}\label{hybr2}
  S_\mathrm{comb}^\opt \cong \frac12 - \frac{\Gamma'^2/2}{\Gamma_\omega + \Gamma_\kappa + 2\sqrt{\Gamma_\omega \Gamma_\kappa + \Gamma'^2 \gamma_m^2}}.
\end{equation}

\begin{figure}[htpb]
 \includegraphics[width=0.5\textwidth]{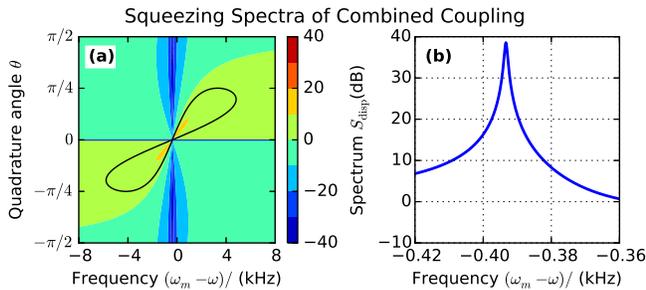}
 \caption{\label{fig1}{(Color online) The density plots (a) and the optimal squeezing quadrature (b) of the output field spectra with combined dispersive and dissipative couplings. The regions inside the black contours in (a) show the $3$dB squeezing region and the blue horizontal line shows the optimal quadrature which is plotted in (b). The coupling strengthes are $G_\omega=2\pi\times75$kHz and  $G_\kappa=2\pi\times15$kHz with driving laser power $\mathcal{P}\sim 40$mW.  Other parameters are identical to those used in Fig.~\ref{fig3}. }}
\end{figure}

In Fig.~\ref{fig1}(a), we plot the output spectra at different quadratures when the optomechanical system is subject to both dispersive and dissipative couplings. We set the coupling strengths such that $G_\omega=5G_\kappa$ in accordance with the experiment parameters in~\cite{combined}.  The density plot resembles the main feature of ponderomotive squeezing with purely $G_\omega$ or $G_\kappa$, except for a trivial quadrature difference. However, there are distinctions. The frequency bandwidth of the squeezing spectra increases at large quadrature angle and shrinks at lower quadrature angles. This is particularly advantageous in practice, since one usually focuses on a specific quadrature and hence one can make use of the larger bandwidth of the squeezed spectra.

In the optomechanical ponderomotive squeezing process, the mechanical element functions as an active mediating element and it provides coherent coupling between two field quadratures. At the same time, it is subject to the environmental Brownian noise which is incoherent with the cavity field. In the reported ponderomotive squeezing experiments with purely dispersive optomechanical coupling, the environment temperature sets the limit of the squeezing magnitudes: Safavi-Naeini \etal~\cite{sq3} reported $0.2$dB squeezing at $\bar{n}_\tth\sim10^4$ and Purdy \etal~\cite{PRX} pushed the squeezing magnitude to $1.7$dB with a lower thermal phonon occupancy $\bar{n}_\tth=47$.
\begin{figure}[t]
 \includegraphics[width=0.5\textwidth]{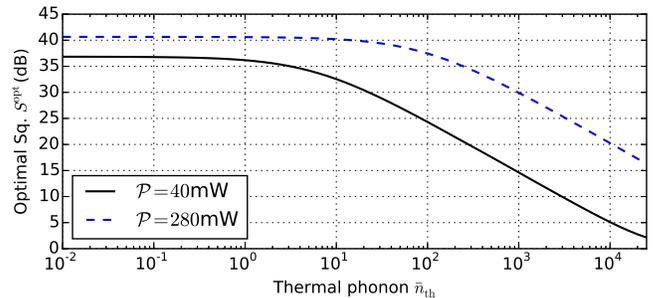}
 \caption{\label{fig4}{(Color online) The effects of the mean thermal phonon occupation $\bar{n}_\tth$ on the optimal squeezing magnitudes with different couplings. The optimal squeezing magnitudes are very similar for finite $\bar{n}_\tth$; hence the three curves overlap.  }}
\end{figure}

We now compare the effect of the thermal phonons on squeezing with different optomechanical couplings. Equations.~(\ref{diss2}), (\ref{disp2}) and (\ref{hybr2}) indicate that the output quadrature variance increases approximately proportionally to $\bar{n}_\tth$ at large coupling rates. Comparing Eqs.~(\ref{diss2}) and (\ref{disp2}), we find that optomechanical systems with purely dissipative coupling ($G_\kappa$) or purely dispersive coupling ($G_\omega$) can generate squeezed field of similar squeezing magnitude if $G_\kappa\omega_m= \sqrt{2}G_\omega\kappa_s$. In Fig.~\ref{fig4}, we illustrate the effects of the mean thermal phonon number on the optimal squeezing magnitude under different coupling regimes. The curves show that the squeezing magnitudes decreases with large thermal phonon occupancy $\bar{n}_\tth$. Even when the thermal phonon number is as high as $\bar{n}_\tth=1000$, the system yields about $15$dB squeezing with combination optomechanical couplings at $\mathcal{P}=40$mW. If we increase the driving laser power to $\mathcal{P}=280$mW, the squeezing magnitude increases to $30$dB. Note that this phonon number is however difficult to achieve with low mechanical frequency $\omega_m$ since $\bar{n}_\tth$ is inversely proportional to $\omega_m$. For example, the system has to be pre-cooled down to $T\sim 6.5$mK in order to get $\bar{n}_\tth=1000$. On the other hand at high bath temperature, a large squeezing magnitude requires an increase in the coupling strength, which can be achieved by increasing the pump power. If the bath temperature increases to $T=1$K, the corresponding thermal phonon number increases to $\bar{n}_\tth \sim 1.5\times10^5$. One needs to increase the driving laser power to $\mathcal{P}\sim280$mW in order to get $3$dB squeezing. For this power the system is still in the stable region.

Note however that the pump power cannot increase infinitely as too strong a pump laser leads to instability of the system dynamics. We discuss the stability condition in detail using the Routh-Hurwitz criterion~\cite{Hurwitz} in the Appendix. For example, our linearization method breaks down and the system settles into instability when the laser power reaches $\mathcal{P}\sim290$mW for the parameters given above and the laser frequency set as  $\omega_l=\omega_0 -\omega_m$. At this power, the coupling strengths are $G_\omega=2\pi\times200$kHz and $G_\kappa=2\pi\times40$kHz. A lower driving laser frequency allows for a higher critical pump power. Our analysis also reveals an unstable region when the effective driving laser detuning $\Delta_s$ is a small negative value. This explains the special instability region discovered in~\cite{combined}.

\section{Squeezing with a fixed frequency driving laser}\label{fixed}
Sawadsky \etal~\cite{combined} demonstrated a strong cooling effect in an optomechanical system with both dissipative and dispersive coupling interactions. The experimental results agree remarkably well with the theoretical calculation. In the experiment, the authors fixed the driving laser frequency $\omega_l$ on resonance with the empty cavity resonance frequency $\omega_c$.
\begin{figure}[phtb]
 \includegraphics[width=0.5\textwidth]{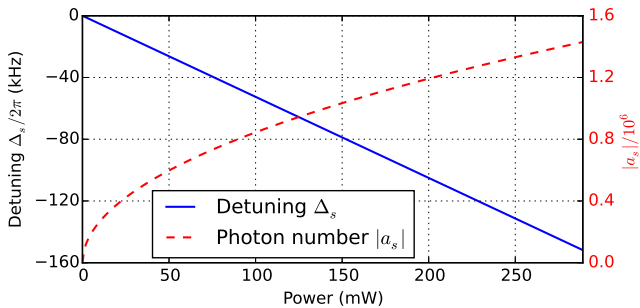}
 \caption{\label{fig5}{(Color online) The change of the effective detuning and mean cavity photon number as the driving laser power increases from $0$ to $200$mW. Other parameters are identical to those used in Fig.~\ref{fig1}.}}
\end{figure}
When the driving laser power increases, the effective cavity resonance frequency changes due to the displacement of the mechanical membrane and this leads to an effective detuning of the driving laser. In this section, we analyze the squeezing phenomena in an optomechanical system driven by a laser with fixed frequency $\omega_l=\omega_c$. Under this condition, the effective detuning $\Delta_s$ and effective cavity decay rate $\kappa_s$ can be determined by solving the nonlinear equation set (\ref{steady}). We use the parameters reported in~\cite{combined}. The solution to (\ref{steady}) shows that $\kappa_s \sim \kappa$ when the driving laser power $\mathcal{E}_l$ is below $250$mW. However, the effective driving laser detuning $\Delta_s$ increases linearly from $0$ to a value close to $-\omega_m$, as shown in Fig.~\ref{fig5}. The cavity mean photon number $|a_s|$ is also displayed in Fig.~\ref{fig5}. When the driving laser power is set as $40$mW, the effective detuning $\Delta_s=2\pi\times 20$kHz. The corresponding coupling strengths remain the values $G_\omega=2\pi\times75$kHz, which are similar to the ones used in Figs.~\ref{fig3} and \ref{fig1}. 
We show the squeezing spectra with combined coupling interaction in Fig.~\ref{fig6} at zero temperature. The optimal squeezing magnitude reaches close to $40$dB. We find large regions with over $3$dB squeezing in the spectrum, as illustrated between the thick black $3$dB contour lines. We observe large regions of squeezing over $10$dB and even squeezing over $20$dB in Fig.~\ref{fig6}. The result is very similar to the one in Fig.~\ref{fig1}(a), and even the effect of temperature is similar so it is not discussed here.

\begin{figure}[t]
 \includegraphics[width=0.5\textwidth]{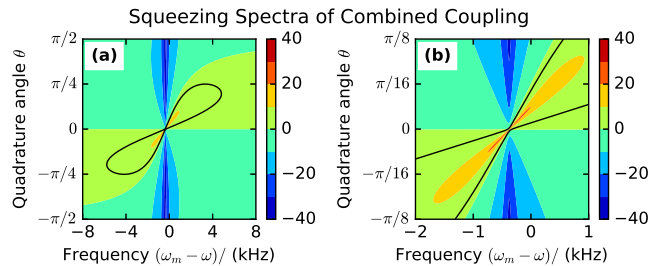}
 \caption{\label{fig6}{(Color online) The squeezing spectra in an optomechanical system with both couplings combined with $G_\omega=5G_\kappa=2\pi\times75$kHz and $\mathcal{P}\sim 40$mW. Different ranges of frequencies and quadrature angles are shown in (a) and (b). Other parameters are identical to those used in Figs.~\ref{fig3} and \ref{fig1}. The regions between the black contours have over $3$dB squeezing. }}
\end{figure}

\section{Conclusion}\label{conclusion}
In conclusion, we investigated the generation of quadrature squeezed states with dissipative coupling optomechanical interactions. Our results show that the dissipative coupling interaction is able to generate strong squeezed vacuum states. The squeezing magnitude depends on the coupling strengths and the mean phonon occupancy due to the mechanical noise. When the dissipative and dispersive coupling strengths are similar, they both generate comparable squeezing magnitude. This scheme works in the unresolved-sideband limit which enables its application in low-frequency mechanical oscillators.
In potential experimental realizations, one challenge would be the large thermal noise introduced by the large phonon number with low mechanical frequency. Large squeezing magnitudes require one to precool the system using a dilution refrigerator. The large thermal noise can also be offset by increasing the pump power by the same order of magnitude as for $\bar{n}_\tth$.

\section*{Appendix: Stability Criterion}

While we follow the standard linearization procedure in solving the nonlinear Hamiltonian, we must make sure the stability of the system dynamics for our chosen parameters.  We investigate the dynamics of the system using the quantum Langevin equation
\begin{equation}\label{14}
  \md\Psi(t)/\md t = \mathbb{M}\Psi(t) + \Psi_\tin(t),
\end{equation}
with $\Psi(t) = (X, Y, Q_1, P_1)^T$ for the system operators, $\Psi_\tin(t) = (\sqrt{2\kappa_s}X_\tin, \sqrt{2\kappa_s}Y_\tin, 0, \xi + \frac{\im G_\kappa}{\sqrt{2\kappa_s}}X_\tin - \frac{\re G_\kappa}{\sqrt{2\kappa_s}}Y_\tin)^T$ for the input noises, and
\begin{widetext}
\begin{equation}\label{15}
  \mathbb{M} \cong \left(
                 \begin{array}{cccc}
                   -\kappa_s & \Delta_s & \re[G_\kappa] - \im[G_\omega] - \frac{g_\kappa}{\sqrt{2\kappa_s}}\mathcal{E}_l & 0 \\
                   -\Delta_s & -\kappa_s & \im[G_\kappa] + \re[G_\omega] & 0 \\
                   0 & 0 & 0 & \omega_m \\
                   \re[G_\omega] & \im[G_\omega] + \frac{g_\kappa}{\sqrt{2\kappa_s}}\mathcal{E}_l & -\omega_m & -\gamma_m \\
                 \end{array}
               \right).
\end{equation}
\end{widetext}
Note $G_{\omega,\kappa} = \frac{2 \sqrt{\kappa_s}\mathcal{E}_l}{\kappa_s + i \Delta_s}$. 
The system is stable if all the eigenvalues of the matrix $\mathbb{M}$ have negative real parts.  This can be examined by applying the Routh-Hurwitz criterion to the polynomial of its eigenvalues. The Routh-Hurwitz coefficients are 
\begin{widetext}
\begin{equation}
\begin{aligned}
h_1 &= 2\kappa_s + \gamma_m, \\
h_2 &= 2 \gamma_m\kappa_s +\Delta_s ^2+\kappa_s ^2+\omega_m^2, \\
h_3 &= \gamma_m\left(\Delta_s ^2+\kappa_s ^2\right) + 2 \kappa_s  \omega_m^2  +\mathcal{E}_l ^2 g_\kappa\omega_m \left(\frac{\sqrt{2} g_\kappa\Delta_s - 4 g_\omega \kappa_s }{\Delta_s ^2+\kappa_s ^2}\right), \\
h_4 &=  \omega_m^2 \left(\Delta_s ^2+\kappa_s ^2\right)-\Delta_s  \mathcal{E}_l ^2 g_\kappa^2\frac{\omega_m}{2\kappa_s}  + \frac{2\sqrt{2}\mathcal{E}_l ^2 \omega_m }{\kappa_s^2 + \Delta_s^2 } (g_\kappa\kappa_s +\Delta_s g_\omega ) (g_\kappa \Delta_s - \sqrt{2}g_\omega \kappa_s). 
\end{aligned}
\end{equation}
\end{widetext}
Routh-Hurwitz criterion reads that the system is stable if and only if all the coefficients $h_i >0$ for $i=1,2,3,4$ and the determinants of all of the Hurwitz matrices are positive, \ie 
\begin{align*}
h_3 &> 0, \\
h_4 &> 0, \\
h_1 h_2 - h_3 &> 0 , \\
h_1 h_2 h_3 - h_1^2 h_4 - h_3^2 &> 0. 
\end{align*}

Note that when $g_\kappa\to0$, these conditions reduce to the stability condition for the optomechanical system with purely dispersive coupling $g_\omega$.

\end{document}